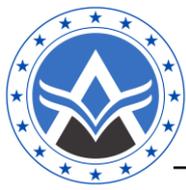

# Interpreted Investigation Report: Loss of Vikram Lander During Lunar Landing Phase

Malaya Kumar Biswal M∗

*Acceleron Aerospace, Bangalore, Karnataka, India – 560037*
**ORCID: 0000-0002-0181-8125**

**Abstract:** This article examines India's first science lander mission on 22 July 2019, attempting a historic landing on the Lunar South Pole Region. Communication was lost at 2.1 km above the lunar surface during the rough braking phase. The cause of the Chandrayaan 2 lander "Vikram" failure remains undisclosed. Possible factors such as vibrations, thruster issues, and power depletion are considered. Recommendations include backup power sources and direct communication systems for interplanetary missions. Despite the setback, ISRO proposed "Chandrayaan 3" to explore the lunar polar region. Chandrayaan 2's legacy influences future missions, shaping India's aspirations for pioneering space endeavors. Gratitude is expressed to ISRO for insights gained during live coverage.

## Table of Contents

## 1. Introduction

On 22 July 2019, India embarked on its pioneering science lander mission by launching the GSLV. The mission was ambitious, aiming to achieve the first-ever landing on the Lunar South Pole Region. Unfortunately, before reaching its intended landing site, the mission encountered a communication failure at an altitude of 2.1 km above the lunar surface during the rough braking of the landing phase. Despite subsequent efforts to reestablish contact with the lander, the mission ultimately remained unsuccessful. The failure of the Chandrayaan 2 lander "Vikram" and its accompanying micro-rover "Pragyaan" has yet to be officially disclosed by ISRO. In this paper, we explore potential causes responsible for the loss of the lander and the mishap that followed rocket.

## 2. Mission Overview

The Chandrayaan 2 mission was initiated on July 22, 2019, at 21:21 UTC through the launch of the Geosynchronous Satellite Launch Vehicle (GSLV) from the Satish Dhawan Space Centre (SHAR). The mission included an orbiter and the Vikram lander. On August 20, 2019, the orbiter and Vikram lander approached the Moon, and subsequent to this, the spacecraft executed its first Lunar bound maneuver, resulting in a successful capture by the Moon's orbit **[2-3]**.

After the successful orbital insertion, the spacecraft executed five additional lunar bound maneuvers. These maneuvers were aimed at facilitating the separation of the Vikram lander from the main bus orbiter, a critical step leading up to the lunar landing attempt. On September 6, 2019, at 20:08 UTC, the Vikram lander was separated from the main orbiter and initiated its descent for landing. Initially, all mission parameters remained within the expected range as the Vikram lander descended to an altitude of over 2.1 km above the lunar surface. However,

---

upon reaching this altitude, at 20:23 UTC (approximately 15 minutes into the landing phase), communication with the lander was unexpectedly lost. This communication failure ultimately resulted in the failure of the lunar landing attempt [2-3].

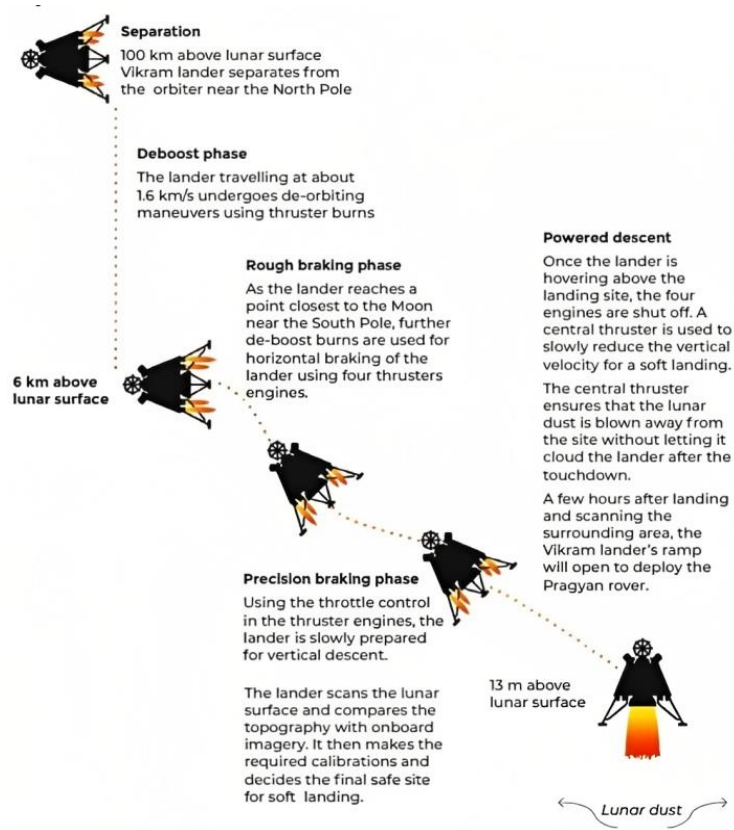

Figure 1 Vikram Lander's Landing Sequence [Courtesy: ISRO]

The detailed sequence of the Vikram lander's landing process is depicted in Figure 1. Subsequent to the landing attempt, efforts were undertaken to locate the impact site of the lander. It was Shanmuga Subramanian who first reported notable changes in lunar surface activity, ultimately aiding in the identification of the lander and its impact site by NASA. The crash site of the lander was determined to be at coordinates 70.8810°S, 22.7840°E, with an elevation of 834 meters (Source: NASA), as illustrated in Figure 3 [1].

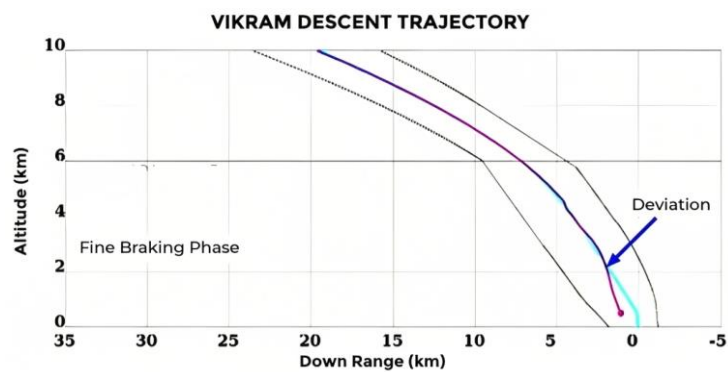

Figure 2 Vikram Lander's Trajectory [Courtesy: ISRO]

3. Interpreted Causes for the Failure of Vikram Lander

- During the phases of rough braking and powered descent, the intense vibrations generated could potentially have dislodged or disengaged the power cord of the communication system (or antenna), resulting in the subsequent loss of communication with the ground.

- At the juncture where a trajectory graph indicates a deviation, it's conceivable that the cessation of thruster activity or incorrect firing of descent thrusts might have led to the spacecraft's loss of control over its orientation. This, in turn, could have resulted in the crash of the lander onto the lunar surface.
- Throughout the descent phase, no signs of spacecraft power were evident. Consequently, the exhaustion of power—either from the onboard battery or solar panels—may have caused the spacecraft's computer to shut down, leading to the simultaneous deactivation of thrusters and the loss of communication. Notably, the lander's programming dictates an automatic landing procedure in the event of interrupted communication issues. In such a scenario, the lander would have executed a safe and soft lunar landing. However, the observed crash of the lander indicates power depletion as a contributing factor during the descent phase.
- The depletion of spacecraft power could potentially be attributed to the limited availability or minimal solar irradiance present in the southern polar region of the Moon [4-10].

### 4. Actual Cause for the Failure of Vikram Lander

During the phase known as Camera Coasting, ISRO took measures to stabilize the altitude of the Vikram Lander at around 7-8 kilometers. This involved fixing both horizontal and vertical orientations in a single direction. However, although the Vikram lander was able to manage its horizontal velocity, it experienced a loss of control over its vertical velocity, leading to a downward acceleration. As the lander transitioned into the rough braking phase, the engine thrust unexpectedly reached maximum levels. This excessive thrust aggravated the inability to control the lander's vertical velocity, resulting in the generation of erroneous trajectory data. Consequently, this erroneous data led to a trajectory deviation, as illustrated in Figure 2. Furthermore, the lander's designed with an orientation adjustment of 10 degrees per second, but it could not be maintained. This led to a loss of control over the lander's orientation, culminating in a crash landing at an unspecified location (Refer Figure 3). In addition to these challenges, the lander encountered a timing issue where the timer experienced a brief interruption, reducing the effective landing duration. This timing anomaly further contributed to the unsuccessful landing attempt. Subsequent to these malfunctions and the subsequent necessary corrections, Chandrayaan 3 successfully addressed these issues, achieving a successful landing on the lunar surface in the South Polar Region on August 23, 2023 [7-10] [Dr.Somnath (ISRO) explained about the root cause at IISc, Bangalore].

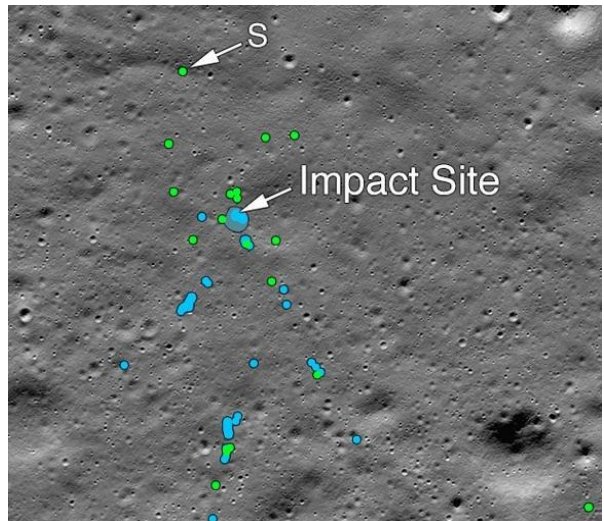

**Figure 3 Impact Site of Vikram Lander [Credit: LRO, NASA]**
**(Color Code: Green – Debris; Blue – Soil Disturbance S – Debris found by Subramanian)**

### 5. Recommendations

Based on the comprehensive analysis conducted following the incident involving the Vikram Lander, we extend our key recommendations to the broader space industry. Specifically, we advise space organizations to consider the integration of Nuclear Thermoelectric Generators (NTG) or Radioisotope Thermoelectric Generators (RTG) as secondary power sources, serving as reliable backups for spacecraft engaged in missions to destinations such as the South Lunar Region or even locations beyond the Mars orbit. This strategic implementation is motivated by the challenges associated with solar irradiance and sunlight availability, which have been identified as primary factors contributing to power loss and battery depletion [8].

Furthermore, we advocate for the adoption of a direct communication framework for both lander and rover missions targeting interplanetary destinations. By employing this approach, the potential impact of communication system impairments can be mitigated. In the event that one communication system encounters difficulties, the alternative channel would ensure continuous and robust connectivity. This dual communication system setup not only to maintain mission reliability but also guarantees the ability to promptly ascertain mission status even in the face of unforeseen communication disruptions. These recommendations, derived from a comprehensive analysis of the Vikram Lander's incident, offer valuable insights into enhancing mission success rates and fortifying communication integrity for ventures aimed at celestial bodies across the cosmos.

## 6. Conclusion

Following the setback encountered during the initial lander mission, ISRO has taken a resilient stance by proposing a subsequent landing demonstration mission known as "Chandrayaan 3." This upcoming endeavor will encompass both a lander and a rover, signifying a steadfast commitment to advancing lunar exploration capabilities. Marking the third instance of India's polar lunar exploration missions, Chandrayaan 3 is a testament to ISRO's perseverance in the realm of space exploration. Derived from the invaluable insights gleaned from the preceding "Vikram" lander mission, Chandrayaan 3 embodies a thorough incorporation of lessons learned. The mission design has been meticulously refined, encompassing various enhancements and refinements. Its primary objective revolves around showcasing proficient landing capabilities tailored for the exploration of the Moon's polar region.

Notably, Chandrayaan 3 underscores an international collaboration with Japan. In this collaborative pursuit, ISRO takes on the role of providing the lander, while the Japanese Aerospace Exploration Agency contributes both the launcher and a rover to the mission. Diverging from its predecessor, Chandrayaan 3 introduces novel components, such as the incorporation of a Laser Doppler Velocimeter (LDV) and cutting-edge night survival technologies. These additions hold immense promise for facilitating site sampling exploration in a more comprehensive and innovative manner. In essence, Chandrayaan 3 stands as a testament to ISRO's unyielding commitment to space exploration, integration of learned lessons, and strategic partnerships in pursuit of pioneering achievements in lunar exploration.

## 7. Legacy of Chandrayaan 2 and Future Prospect

The name "Vikram" assigned to the lander reverentially pays homage to the founding father of the Indian Space Program, Dr. Vikram Sarabhai. This mission heralded a monumental achievement for India, marking the nation's inaugural endeavor to land on an interplanetary destination. Despite the challenges faced during its execution, the trajectory data and technological lessons garnered from the Vikram lander's demonstration stand as a guiding light for forthcoming lander missions. These forthcoming ventures encompass not only Mars but also the wider expanse beyond.

As we look to the future, the aspirations extend to Mars Orbiter Mission 2 (Mangalyaan 2), projected to materialize within the next half-decade. This ambitious endeavor potentially involves the incorporation of both a lander and a rover. The journey set forth by Chandrayaan 2 has effectively paved the way for such audacious initiatives. By capitalizing on the acquired knowledge, India's space endeavors are poised to embrace Mars and beyond with enhanced capability and renewed vigor [11-12].

## 8. Note & Report Statement

The conclusions and insights presented here are interpreted from a comprehensive examination of lunar and Mars probe failures, offering interpretations that aren't conclusive about the precise triggers behind the lander mishap. It's important to note that ISRO has yet to disclose the official investigation report detailing the circumstances surrounding the Vikram lander incident. As a result, this study has been undertaken to shed light on the potential factors that might have contributed to the setback faced by the Indian lander, "Vikram."

Given the broader context, it's imperative to recognize the significance of the failure report in shaping future prospects. In light of this, we've delved into the conceivable causes that could have played a role in the Chandrayaan 2 lander's unsuccessful outcome. It's crucial to underscore that these speculations are an attempt to fill the knowledge gap in the absence of the official investigation findings and should be treated as such.

## 10. Biography

Malaya Kumar Biswal is the Founder & CEO of Acceleron Aerospace in Bangalore, India. He is a renowned entrepreneur and scientist in aerospace engineering, specializing in space exploration. After earning his bachelor's degree in Aerospace Engineering, he worked as a Senior Research Scientist at Grahaa Space, focusing on satellite reliability, aerospace design, and space science research. With a visionary mindset, Biswal founded Acceleron Aerospace and now leads the company in revolutionizing the aerospace industry. He is particularly interested in human Mars exploration and envisions ambitious missions to Mars and Ceres. Biswal's achievements have earned him respect in the scientific community, and he actively mentors aspiring scientists, inspiring future space pioneers.

## 11. Acknowledgement

## 12. Conflict of Interest

The author have no conflict of interest to report.

## 13. Funding

No external funding was received to support this study.